\documentclass[prd,tightenlines,twocolumn]{revtex4}
\usepackage{amsmath}
\usepackage{amssymb}
\usepackage{graphicx}
\usepackage{bm}
\usepackage{enumerate}
\usepackage{color}
\newcommand{\be}{\begin{equation}}
\newcommand{\ee}{\end{equation}}
\newcommand{\bea}{\begin{eqnarray}}
\newcommand{\eea}{\end{eqnarray}}
\newcommand{\ba}{\begin{eqnarray}}
\newcommand{\ea}{\end{eqnarray}}

\begin{document}

\title{Comments on the chemical and kinetic equilibration \\
in heavy ion collisions}
\author{ 
 Edward Shuryak }

\affiliation{Department of Physics and Astronomy, \\ Stony Brook University,\\
Stony Brook, NY 11794, USA}


\begin{abstract} 
I argue that perturbative scattering of quarks and gluons are incompatible with lattice and heavy ion data on QGP properties. 
The non-perturbative mechanisms for quasiparticle rescattering and quark production
are briefly discussed, as well as experiments needed to measure matter anisotropy and
quark density at early stages of the collisions.
 \end{abstract}

\maketitle

\section{Equilibration in weak coupling}
Let me start with my ``naive weak coupling" approach in a quater-century-old paper   
\cite{Shuryak:1992wc}. It was based on the lowest order cross sections
\be {d\sigma(gg\rightarrow gg)  \over dt}=
 \big({\pi \alpha_s^2 \over s^2}\big)  \big({9\over 2}\big) \big(3-{u t \over s^2} - {u s \over t^2 } - {s t \over u^2}\big) \ee
\be {d\sigma(gg\rightarrow \bar q q)  \over dt}= \big({\pi \alpha_s^2 \over s^2}\big)
\big( {u^2+t^2 \over 6 ut}- {3u^2+3t^2\over 8 s^2}\big) \ee
I then noticed that at large angles the ratio of the two is very large $(gg\rightarrow gg)/(gg\rightarrow \bar q q) =  30/0.14\sim 200$. At small angles also, the former has $1/t^2$ 
term while the second has only $1/ut$ term. I then concluded that kinetic equilibration
of glue happens quicker than chemical equilibration of quarks, known as the ``hot glue scenario". 

Before proceeding to more recent works, let me go directly to
the point.
 Note that
in kinetic equilibration of glue one needs to evaluate the transport cross section,
which is divergent logarithmically $\sigma_T= \int (1-cos(\theta)) d\sigma  $. In chemical equilibration
it is the total cross section which matters,  $\sigma=\int d\sigma $, which is   divergent logarithmically as well. 

The proper regulators of the $t$-channel gluon propagator are different for electric and magnetic exchanges. The former is regulated by the so called electric screening mass, which in weak coupling is   \cite{Shuryak:1977ut} 
\be M_E^2=g^2 T^2(1+N_f/6) \label{eqn_M_E}\ee while, as also shown there,  the magnetic fields remains unscreened in pQCD. Furthermore, the magnetic part has structure $$ {1 \over Q^2} \Pi_\perp(Q)  {1 \over Q^2}\sim  {1 \over Q^2} $$
because magnetic polarization tensor is $\Pi_\perp(Q)\sim Q^2$. The combination of electric and magnetic effects leads to the following substitution in the cross section
$$ \big({1 \over Q^2}\big)^2\rightarrow {1\over Q^2(Q^2+M_E^2)} $$
which makes the transport cross section converegent. 

%

\begin{figure}[h!]
\begin{center}
\includegraphics[width=6cm]{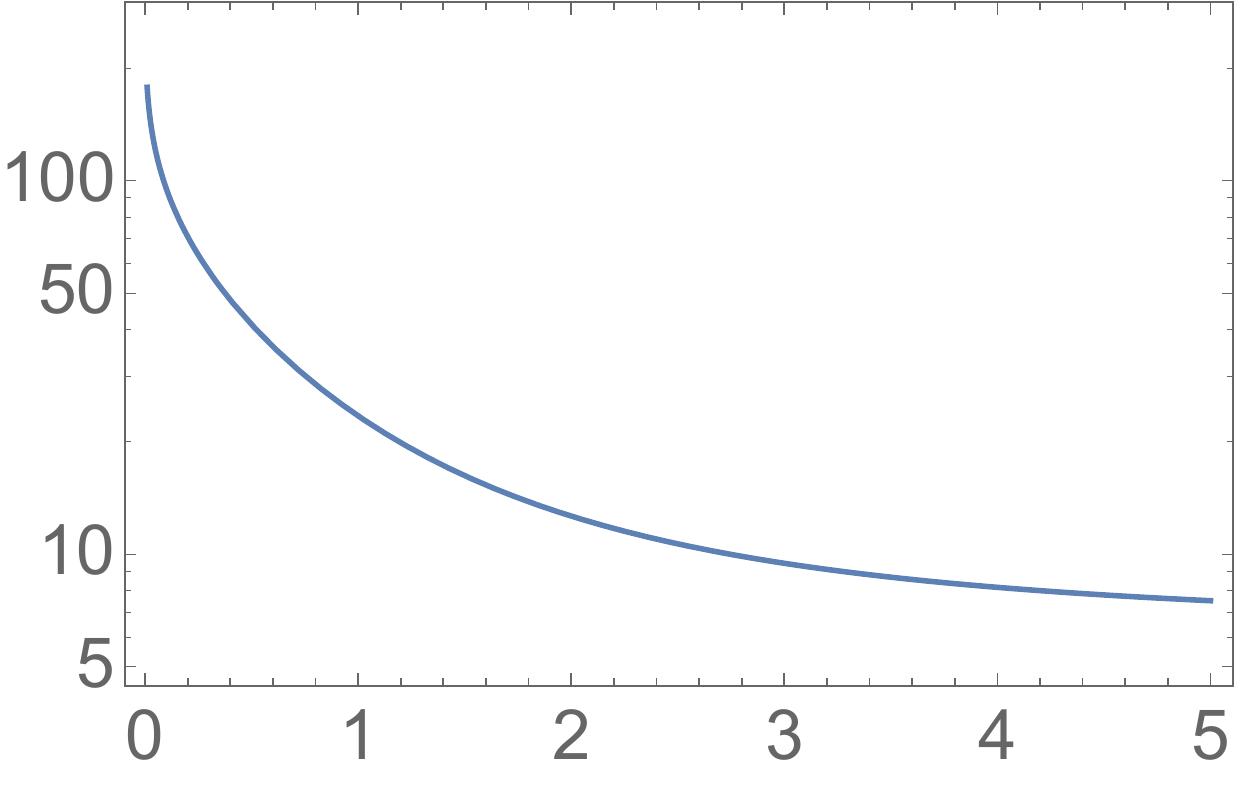}
\caption{The angle-integrated matrix element squared $\int (1-cos(\theta))|M_{gg}|^2 dcos(\theta)$ versus the electric screening mass $M_E/p$ normalized to the particle CM momenta.}
\label{fig_trans_cross}
\end{center}
\end{figure}

Modern version of the kinetics at weak coupling, with consistent IR resummation of relevant higher-order diagrams, was developed in \cite{Arnold:2000dr}, and  issue of quark chemical equilibration
 has been recently discussed in 
\cite{Kurkela:2018oqw}. 
The key role in it is played by gluon exchanges  with  soft 
``splitting" of gluons,  $g\rightarrow gg, \bar q q$. In order to show their role better, we show in Fig\ref{fig_trans_cross} the
squared matrix element ({\em without coupling in front}) integrated over angle to the transport cross
section,  as a function of the electric screening mass (normalized to particle CM momenta $p$).
One can see that, even after we factor out the coupling, there are basically two regimes:
(i)  small screening masses $m_E/p \ll 1$, and (ii) large ones $m_E/p \sim 1$. 
It is the former regime (used in the above-mentioned works) the transport cross seciton
is larger than in the latter by about an order of magnitude.

The main conclusion of Ref.\cite{Kurkela:2018oqw} is that 
there is no time hierarchy of processes discussed, and thus no ``hot glue" scenario: {\em the kinetic and chemical
equilibration happen at the same time.}
 
Their calculation is done for the range of
 the t' Hooft coupling $\lambda\equiv g^2 N_c=0.1, 1,10$ or $\alpha_s\approx 1/300,1/30,1/3$.
The authors of course know that those values do not correspond to realistic
coupling in experimentally produced QGP, and that at such couplings the
both equilibration times are way too long to explain the data, such as collective flows.
Yet they hope one can use 
these calculations to understand the dependence on the coupling of physical quantities and then safely extrapolate their results to ``realistic" couplings.

\section{Can one successfully extrapolate, from weak to strong coupling?}

Let me on the onset say that my theoretical prejudice is to answer this question $negatively$.
Let me outline two general theoretical reasons for this opinion, before plunging into details.

(i) In weak coupling the matter is in a gas-like phase, and therefore particle interactions are adequately represented by a kinetic equation, with a {\em mean free path} as a key parameter. 
The famous Boltzmann's hypothesis, that manybody distributions all factorize into a product of single-body ones,
is justified.  However, when coupling is large and potential energy is comparable to temperature, the
matter changes to a liquid-like form, and eventually solidifies. Two- (and more) particle correlations are present
permanently, there is no Boltzmann reduction 
 and cascades.  There are  no $in$ and $out$ states, with particles not going to infinity but being near each other all the time.   It is well documented even for many classical systems,
 studied by molecular dynamics.
 
(ii) The QCD-like non-Abelian theories and at $T/T_c\sim O(1)$ are known to posess strong 
non-perturbative phenomena, induced by gauge fields in forms of solitons -- monopoles,
instantons, instanton-dyons. Those are invisible in pQCD but play a signifiant role
in dynamics (more of that at the end). 

 Let us for now return to Ref.\cite{Kurkela:2018oqw}. For their three values of the coupling 
  the viscosity-to-entropy-density-ratio $\eta/s$ is equal to 1900, 35 and 1, respectively.
 The last value is about factor 6 from the empirical value, and, taken large spread of values,
 one may think that extrapolation to it may be possible. 
 
  \begin{figure}[htbp]
\begin{center}
\includegraphics[width=6cm]{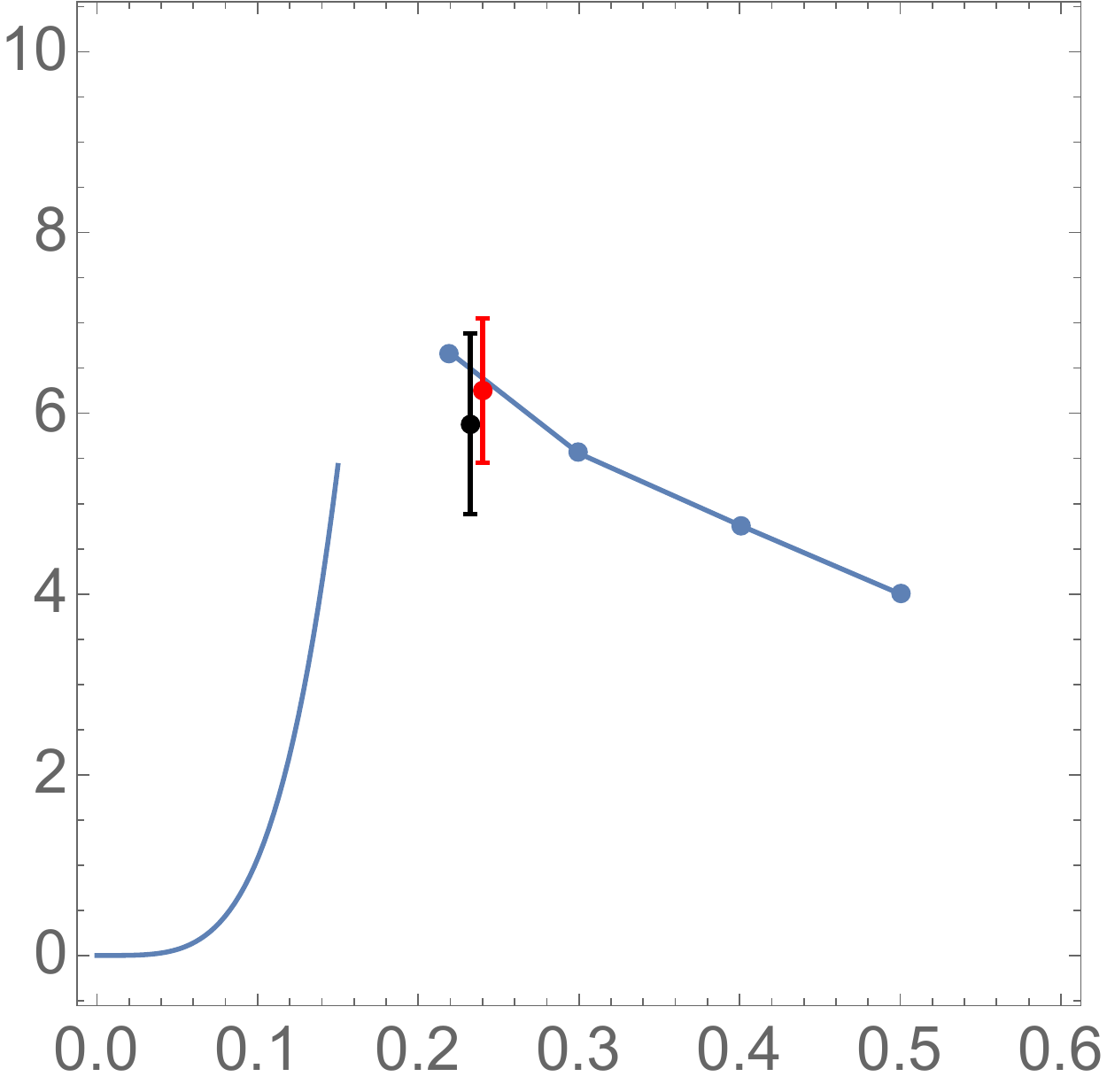}
\includegraphics[width=7cm]{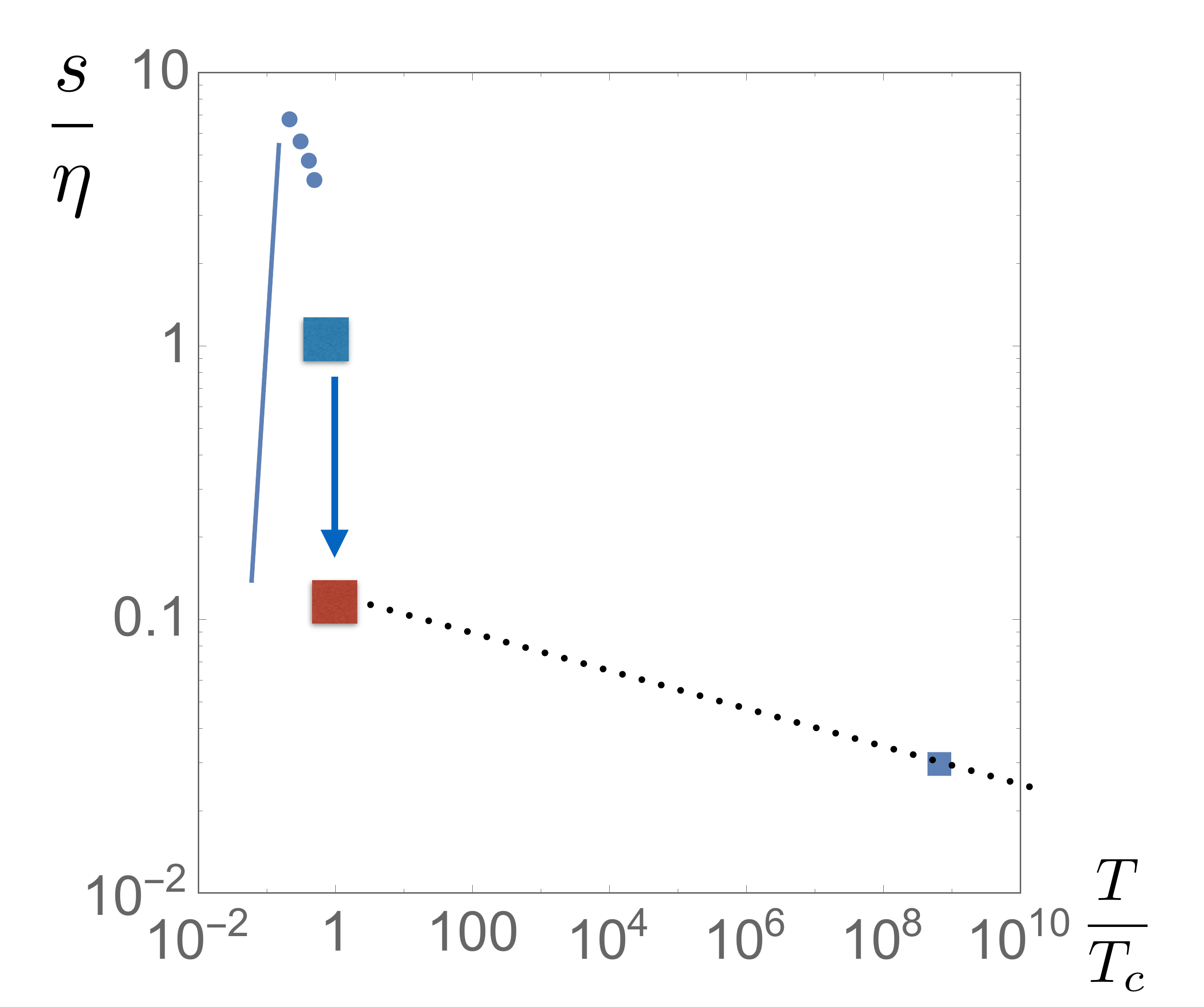}
\caption{Both plots show the inverse $s/\eta$ ratio versus the temperature $T\, (GeV)$. The upper linear plot shows the empirical value (red) and the  lattice result \cite{Nakamura:2004sy} ,with the error bars.
The four points with line are from \cite{Ratti:2008jz}, representing  gluon-monopole scattering.
The line without points on the left corresponds to pion rescattering.
The lower log-log plot includes also  points from weak coupling cascades \cite{Kurkela:2018oqw} shown by two blue squares. The arrow and red square correpond
to transition from small to large angle regime, discussed in the text. }
\label{fig_SoverEta}
\end{center}
\end{figure}
 
 The situation is shown in Fig.\ref{fig_SoverEta}.  We show two of the three points from 
 \cite{Kurkela:2018oqw}, as the last one corresponding to $T\sim 10^{100} \,GeV$ is hard to
 fit even in the log-log plot. My point is that the key left point, with coupling $\alpha_s\sim 1/3$,
 should $not$ be located as it is calculated kinetically and plotted in  Fig.\ref{fig_SoverEta}(b)
 because under this conditions {\em the perturbative expression for the screening mass 
 (\ref{eqn_M_E}) is invalid}.  
 
\begin{figure}[htbp]
\begin{center}
\includegraphics[width=8cm]{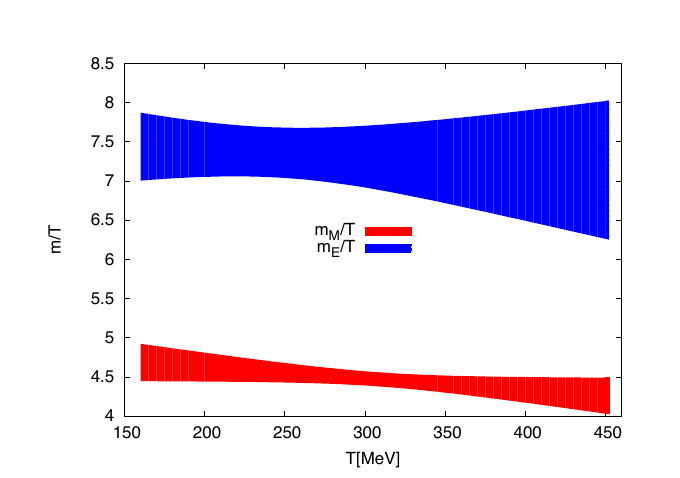}
\caption{The continuum extrapolations of the electric and magnetic screening masses, normalized to the temperature,  from Borsanyi et al }
\label{fig_screening_masses}
\end{center}
\end{figure}

 We know it from lattice studies (which do not require transition
 from Euclidean to Minkowski world, and are thus quite reliable.) 
 In Fig.\ref{fig_screening_masses} one finds relatively recent lattice data, done by well respected Wuppertal-Budapest collaboration and carefully extrapolated to continuum limit, for $M_E/T$ ratio. Instead of being small,
 it is around 7.5 (!).  Other lattice groups give other numbers, but none of them finds a value
 even close to 1, always significantly larger, 5-15. What it means, simply speaking, is that assumed dominance
 of small angle scatterings or ``soft splittings", is in fact in {\em direct contradiction to lattice data}. If one uses the lattice values of the $M_E/T$, one appears in the large angle regime,
 and therefore the transport cross section moves from the left side of Fig.\ref{fig_trans_cross}
 to the right. Then the
 left square point in  Fig.\ref{fig_SoverEta}(b) needs to be moved down, by an
 order of magnitude or so. The second point -- roughly corresponding to 
 coupling in electroweak plasma -- remains at the same place. Now, looking at them,
 one cannot imagine that any smooth  extrapolation to the correct $s/\eta$ value may exist.

 \section{The magnetic screening, monopoles and viscosity} 

Let me give now two more general reasons why one needs to think here about 
magnetic monopoles, and not extrapolation of soft gluon scattering.  

(i)
A shortcoming of pQCD  viewpoint is that it {\em cannot explain a nonzero magnetic screening mass }. 
But, as we see from the lattice results just shown in Fig.\ref{fig_screening_masses}, the magnetic mass is not only nonzero, but is even comparable to the electric one! 

(ii) the peak in $s/\eta$ shown in Fig. \ref{fig_SoverEta}  is located near the 
deconfinement temperature $T_c$, and therefore one may suspect that
one should be related to the other.  
%
The monopoles, detected on the lattice, have density peaking near $T_c$, with the magnitude
  comparable to that of quarks and gluons. Furthermore, this density is large enough for
 their Bose-Einstein condensation (BEC) to occur, at the deconfinement transition $T=T_c$.  Multi-monopole Bose-clusters were observed on the lattice \cite{DAlessandro:2010jdd}
and studied in Path Integral Monte Carlo \cite{Ramamurti:2017fdn}. The dance made by monopoles near $T_c$ is remarkably similar
to that of $^4He$ atoms near the lambda point.

It was shown in \cite{Ratti:2008jz} that gluon scattering on 
these monopoles has interesting angular distribution, with a backward peak.  Unlike $gg$ scattering, it gives the transport cross section consistent
with the observed $\eta/s$. 

More recently, the  puzzling angular  distribution of the jet quenching was explained
by the monopole contribution, see \cite{Xu:2015bbz,Ramamurti:2017zjn} .

\section{Quark production via the instanton/sphaleron mechanism}

Perturbative production of quark pairs is not the only way chemical equilibration of QGP
can proceed. Instanton contribution to inelastic hadronic collisions were discussed in 
\cite{Nowak:2000de}. It was then realized that this process leads to instanton-sphaleron
conversion, with subsequent (over-the barrier or Minkowskian) sphaleron explosion
\cite{Shuryak:1992bt}. Recently, in Glasma model framework, the sphaleron 
quark production has been studied in \cite{Mace:2016svc}, which concluded that
this mechanism is quite effective.

There is a very important distinction between the pQCD (both soft or hard) production 
of quarks, and the sphaleron mechanism. The former produce $left-left$ or $right-right$
polarized pairs, and thus does not produce any {\em chiral imbalance} in the QGP. 
On the contrary, the sphaleron mechanism produces final states like $(\bar u_R u_L)  (\bar d_R d_L)(\bar s_R s_L)$, with 6 units of axial charge per event. Although average is
still zero, fluctuations can create chiral imbalance. 

\section{Summary and discussion}

The main point of these comments is that not only a weak coupling regime
is not occurring in realistic QGP conditions, but it is also impossible to get
its kinetic properties by extrapolation from weak coupling. 

Soft kicks following parton splitting works well in jet quenching, the BDMPS theory etc,
because the momentum is large 
$$|\vec p|\sim10-1000\, GeV \gg Q\sim 1 \, GeV$$ 
The QGP constituents with the typical gluon momenta  $p\sim 3T\sim 1 \, GeV$
are not in this regime. They cannot be softly split into two, just because their effective masses
are also $\sim M_E\sim \, 1 \, GeV$.  Furthermore, 
there are no soft gluon exchanges because the screening masses are that large.

The debate between weak and strong coupling scenarios of heavy ion collisions
is in fact rather old. It
was intense around the year 2000, before the RHIC era. The soft gluon exchanges
were the basis of the so called  
 bottom-up scenario \cite{Baier:2000sb}. Through 1990s nearly all high energy physicists
 were telling us that there is no hope to produce new form of matter in heavy ion collisions, and all
 we will see would be a fireworks of minijets, without any collective effects.
  The predictions of Molnar  (in \cite{Bass:1999zq}), based on exactly 
 this soft 
 gluon cascade, was that $v_2$ should drop down at RHIC. 
 
  Fortunately, this pessimistic
 point of view was  spectacularly overthrown in 
 the first years of RHIC operation. The data  confirmed instead the robust hydro explosion, with hydrodynamics describing it quite accurately. The observed elliptic flow
growth with $p_\perp$ to large values was never reproduced by gluon cascades, even with
huge assumed cross sections (completely incompatible with screening masses).  
 
 Further observation of elliptic flow and higher harmonics has lead to viscosity measurement, giving the value we discussed above. 
The notion of strongly-coupled QGP has prevailed. Using AdS/CFT correspondence
one found good description of rapid convergence to hydro regime, in a time of fraction of 
fm/c. The equilibration mechanism was found to proceed in the {\em opposite direction,}
from UV to IR (top-down scenario) \cite{Lin:2008rw}.

Apparent resurgence of weak coupling methods in the last few years looks quite surprising.
I even heard statements at some meetings of ``hydrodynamics without a fluid".  So, let me
state it again: {\em the bottom-up scenario and weakly coupled cascades
in general were before and still  are completely incompatible with the data on elliptic flow}.
It is of course especially obvious for ``small system", $pA,pp$ in which flows were discovered lately. 

While the issue of weak-versus-strong-coupling-equilibration was, in fact, resolved some time ago,
direct measurements of anisotropy of matter at early time would be desirable.
The proposal how to do so by dilepton polarization  has been made in \cite{Shuryak:2012nf}.

The issue of quark production/equilibration however still requires a lot of work. 
Can we check experimentally which mechanism of QGP chemical equilibration
is in place in real-world heavy ion collisions? 

One way proposed is to use dileptons \cite{Shuryak:1992bt}, more specifically the
so called ``intermediate mass dileprons" (IMD) (between $\phi$ and $\psi$ peaks) produced
early in the collision. If the quark production is delayed, one expects a deficit of
such dileptons in respect to standard calculations assuming fully equilibrated QGP.
To my knowledge no such deficit has been reported in all comparisons
made so far, although the accuracy of
that needs to be further investigated.  

A specific consequence of sphaleron mechanism is chiral imbalance, on event-by-event
basis. This was an important assumption in well known proposal to observe the 
 {\em chiral magnetic effect} (CME). Hopefully, recent RHIC
 run with two isotopes of $A=96$  will clarify the effect of magnetic field and magnitude of the
   CME.  As a consequence, it should be able to establish  
the magnitude of the sphaleron production rate.


\newpage
\begin{thebibliography}{99} 

\bibitem{Shuryak:1992wc} 
  E.~V.~Shuryak,
  ``Two stage equilibration in high-energy heavy ion collisions,''
  Phys.\ Rev.\ Lett.\  {\bf 68}, 3270 (1992).

\bibitem{Shuryak:1977ut} 
  E.~V.~Shuryak,
  ``Theory of Hadronic Plasma,''
  Sov.\ Phys.\ JETP {\bf 47}, 212 (1978)
  [Zh.\ Eksp.\ Teor.\ Fiz.\  {\bf 74}, 408 (1978)].

\bibitem{Arnold:2000dr} 
  P.~B.~Arnold, G.~D.~Moore and L.~G.~Yaffe,
  ``Transport coefficients in high temperature gauge theories. 1. Leading log results,''
  JHEP {\bf 0011}, 001 (2000)
  [hep-ph/0010177].

\bibitem{Kurkela:2018oqw} 
  A.~Kurkela and A.~Mazeliauskas,
  ``Chemical equilibration in weakly coupled QCD,''
  arXiv:1811.03068 [hep-ph].

\bibitem{Baier:2000sb} 
  R.~Baier, A.~H.~Mueller, D.~Schiff and D.~T.~Son,
  ``'Bottom up' thermalization in heavy ion collisions,''
  Phys.\ Lett.\ B {\bf 502}, 51 (2001)
  [hep-ph/0009237].

\bibitem{Nakamura:2004sy} 
  A.~Nakamura and S.~Sakai,
  ``Transport coefficients of gluon plasma,''
  Phys.\ Rev.\ Lett.\  {\bf 94}, 072305 (2005)
  [hep-lat/0406009].

\bibitem{DAlessandro:2010jdd} 
  A.~D'Alessandro, M.~D'Elia and E.~V.~Shuryak,
  ``Thermal Monopole Condensation and Confinement in finite temperature Yang-Mills Theories,''
  Phys.\ Rev.\ D {\bf 81}, 094501 (2010)
  [arXiv:1002.4161 [hep-lat]].


\bibitem{Ramamurti:2017fdn} 
  A.~Ramamurti and E.~Shuryak,
  ``Effective Model of QCD Magnetic Monopoles From Numerical Study of One- and Two-Component Coulomb Quantum Bose Gases,''
  Phys.\ Rev.\ D {\bf 95}, no. 7, 076019 (2017)
  [arXiv:1702.07723 [hep-ph]].

\bibitem{Borsanyi}
S. Borsanyi, Z. Fodor, S. D. Katz, A. Pasztor, K. K. Szabo and C. Torok, Static QQ pair free energy and screening masses from correlators of Polyakov loops: continuum extrapolated lattice results at the QCD physical point, JHEP 1504, 138 (2015), [arXiv:1501.02173[hep- lat]].

\bibitem{Ratti:2008jz} 
  C.~Ratti and E.~Shuryak,
  ``The Role of monopoles in a Gluon Plasma,''
  Phys.\ Rev.\ D {\bf 80}, 034004 (2009)
  [arXiv:0811.4174 [hep-ph]].

\bibitem{Xu:2015bbz} 
  J.~Xu, J.~Liao and M.~Gyulassy,
  ``Bridging Soft-Hard Transport Properties of Quark-Gluon Plasmas with CUJET3.0,''
  JHEP {\bf 1602}, 169 (2016)
  [arXiv:1508.00552 [hep-ph]].

\bibitem{Ramamurti:2017zjn} 
  A.~Ramamurti and E.~Shuryak,
  ``Role of QCD monopoles in jet quenching,''
  Phys.\ Rev.\ D {\bf 97}, no. 1, 016010 (2018)
  [arXiv:1708.04254 [hep-ph]].

\bibitem{Nowak:2000de} 
  M.~A.~Nowak, E.~V.~Shuryak and I.~Zahed,
  ``Instanton induced inelastic collisions in QCD,''
  Phys.\ Rev.\ D {\bf 64}, 034008 (2001)
  [hep-ph/0012232].

\bibitem{Shuryak:2002qz} 
  E.~Shuryak and I.~Zahed,
  ``Prompt quark production by exploding sphalerons,''
  Phys.\ Rev.\ D {\bf 67}, 014006 (2003)
  doi:10.1103/PhysRevD.67.014006
  [hep-ph/0206022].

\bibitem{Mace:2016svc} 
  M.~Mace, S.~Schlichting and R.~Venugopalan,
  ``Off-equilibrium sphaleron transitions in the Glasma,''
  Phys.\ Rev.\ D {\bf 93}, no. 7, 074036 (2016)
  doi:10.1103/PhysRevD.93.074036
  [arXiv:1601.07342 [hep-ph]].

\bibitem{Bass:1999zq} 
  S.~A.~Bass {\it et al.},
  ``Last call for RHIC predictions,''
  Nucl.\ Phys.\ A {\bf 661}, 205 (1999)
  [nucl-th/9907090].

\bibitem{Lin:2008rw} 
  S.~Lin and E.~Shuryak,
  ``Toward the AdS/CFT Gravity Dual for High Energy Collisions. 3. Gravitationally Collapsing Shell and Quasiequilibrium,''
  Phys.\ Rev.\ D {\bf 78}, 125018 (2008)
  [arXiv:0808.0910 [hep-th]].

\bibitem{Shuryak:1992bt} 
  E.~V.~Shuryak and L.~Xiong,
  ``Dilepton and photon production in the 'hot glue' scenario,''
  Phys.\ Rev.\ Lett.\  {\bf 70}, 2241 (1993)
  [hep-ph/9301218].

\bibitem{Shuryak:2012nf} 
  E.~Shuryak,
  ``Monitoring parton equilibration in heavy ion collisions via dilepton polarization,''
  arXiv:1203.1012 [nucl-th].
\end{thebibliography}
\end{document}